\documentclass[runningheads]{llncs}

\usepackage[hidelinks]{hyperref}
\usepackage[utf8]{inputenc}
\usepackage[small]{caption}
\usepackage{graphicx}
\usepackage{amsmath}
\usepackage{makecell}
\usepackage{booktabs}
\usepackage{algorithm}
\usepackage{algorithmic} 
\usepackage[switch]{lineno}
\usepackage{tikz}
\def\checkmark{\tikz\fill[scale=0.4](0,.35) -- (.25,0) -- (1,.7) -- (.25,.15) -- cycle;} 
\DeclareMathOperator*{\argmin}{arg\,min}

\newcommand{\modelname}{NPP}
\usepackage{arydshln}
\usepackage{comment}
\usepackage{subfig}

\begin{document}
\title{Neural Pre-Processing: A Learning Framework
for End-to-end Brain MRI Pre-processing}

\titlerunning{Neural Pre-processing}
%
\author{Xinzi He\inst{1}, Alan Wang\inst{2}, Mert R. Sabuncu\inst{1,2}}
\authorrunning{X. He et al.}
%
\institute{School of Biomedical Engineering, Cornell University \\\email{xh278@cornell.edu}\and School of Electrical and Computer Engineering, Cornell University}
%
\maketitle              
\begin{abstract}
Head MRI pre-processing involves converting raw images to an intensity-normalized, skull-stripped brain in a standard coordinate space. 
In this paper, we propose an end-to-end weakly supervised learning approach, called Neural Pre-processing (\modelname), for solving all three sub-tasks simultaneously via a neural network, trained on a large dataset without individual sub-task supervision. 
Because the overall objective is highly under-constrained, we explicitly disentangle geometric-preserving intensity mapping (skull-stripping and intensity normalization) and spatial transformation (spatial normalization). 
Quantitative results show that our model outperforms state-of-the-art methods which tackle only a single sub-task. 
Our ablation experiments demonstrate the importance of the architecture design we chose for \modelname.
Furthermore, \modelname~affords the user the flexibility to control each of these tasks at inference time. 
The code and model are freely-available at \url{https://github.com/Novestars/Neural-Pre-processing}.

\keywords{Neural network \and Pre-processing \and Brain MRI.}
\end{abstract}
\section{Introduction}
Brain magnetic resonance imaging (MRI) is widely-used in clinical practice and neuroscience. 
Many popular toolkits for pre-processing brain MRI scans exist, e.g., FreeSurfer~\cite{fischl2012freesurfer}, FSL~\cite{Smith2004}, AFNI~\cite{cox1996afni}, and ANTs~\cite{Avants2008}. 
These toolkits divide up the pre-processing pipeline into sub-tasks, such as skull-stripping, intensity normalization, and spatial normalization/registration, which often rely on computationally-intensive optimization algorithms. 

Recent works have turned to machine learning-based methods to improve pre-processing efficiency. 
These methods, however, are designed to solve individual sub-tasks, such as SynthStrip~\cite{hoopes2022synthstrip} for skull-stripping and Voxelmorph~\cite{Balakrishnan2019} for registration. 
Learning-based methods have advantages in terms of inference time and performance.
However, solving sub-tasks independently and serially has the drawback that each step's performance depends on the previous step.
In this paper, we propose a neural network-based approach, which we term Neural Pre-Processing (\modelname), to solve three basic tasks of pre-processing simultaneously. 


\modelname~first translates a head MRI scan into a skull-stripped and intensity-normalized brain using a translation module, and then spatially transforms to the standard coordinate space with a spatial transform module. 
As we demonstrate in our experiments, the design of the architecture is critical for solving these tasks together.
Furthermore, \modelname~offers the flexibility to turn on/off different pre-processing steps at inference time.
Our experiments demonstrate that \modelname~achieves state-of-the-art accuracy in all the sub-tasks we consider.

 

\section{Methods}

\subsection{Model}

As shown in Fig.~\ref{fig_p}, our model contains two modules: a geometry-preserving translation module, and a spatial-transform module. 

\subsubsection{Geometry-preserving Translation Module.}
\label{sec:intensity-module} 
\begin{figure}[t] 
\centering
\includegraphics[width=0.88\textwidth]{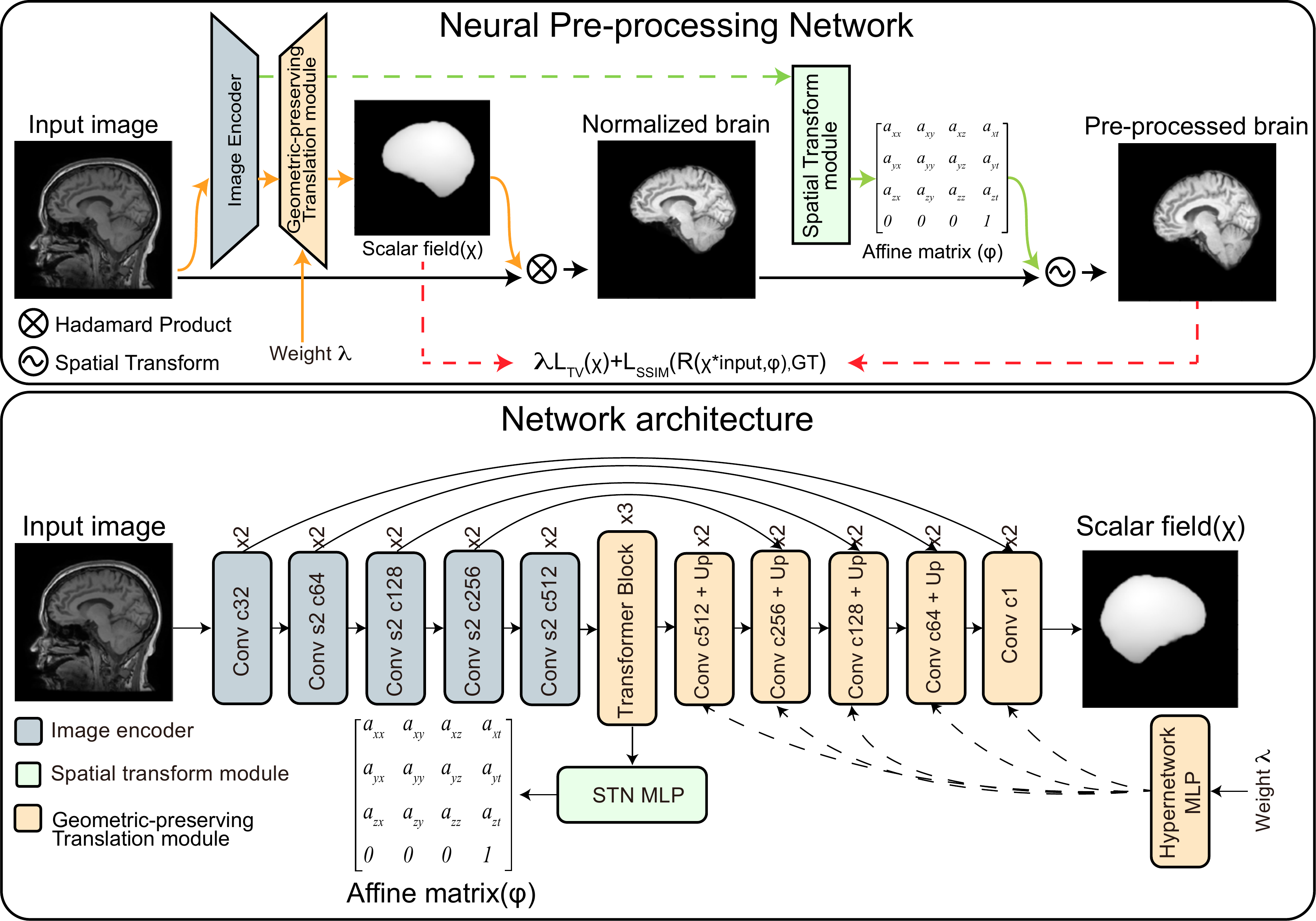} 
\caption{(Top) An overview of Neural Pre-processing. (Bottom) The network archiecture of Neural Pre-processing} \label{fig_p}
\end{figure}

This module converts a brain MRI scan to a skull-stripped and intensity normalized brain. 
We implement it using a U-Net style~\cite{ronneberger2015u} $f_\theta$ architecture (see Fig.~\ref{fig_p}), where $\theta$ denotes the model weights. 
We operationalize skull stripping and intensity normalization as a pixel-wise multiplication of the input image with a scalar multiplier field $\chi$, which is the output of the U-Net $f_\theta$: 
\begin{equation}
    T_{\theta}(x) = f_\theta(x) \otimes x,
\end{equation}
where $\otimes$ denotes the element-wise (Hadamard) product. 

Such a parameterization allows us to impose constraints on $\chi$. In this work, we penalize high-frequencies in $\chi$, via the total variation loss described below.  
Another advantage of $\chi$ is that it can be computed at a lower resolution to boost both training and inference speed, and then up-sampled to the full resolution grid before being multiplied with the input image. 
This is possible because the multiplier $\chi$ is spatially smooth. 
In contrast, if we have $f_\theta$ directly compute the output image, doing this at a lower resolution means we will inevitably lose high frequency information. 
In our experiments, we take advantage of this by having the model output the multiplicative field at a grid size that is 1/2 of the original input grid size along each dimension. 

\subsubsection{Spatial Transformation Module.}
Spatial normalization is implemented as a variant of the Spatial Transformer Network (STN)~\cite{jaderberg2015spatial}; in our implementation, the STN outputs the 12 parameters of an affine matrix $\Phi_{aff}$. 
The STN takes as input the bottleneck features from the image translation network $f_\theta$ and feeds it through a multi-layer perceptron (MLP) that projects the features to a 12-dimensional vector encoding the affine transformation matrix. This affine transformation is in turn applied to the output of the image translation module $T_\theta(x)$ via a differentiable resampling layer~\cite{jaderberg2015spatial}. 

\subsection{Loss Function}
\label{sec:loss}
The objective to minimize is composed of two terms.
The first term is a reconstruction loss $L_{rec}$. 
In this paper, we use SSIM~\cite{Wang2004} for $L_{rec}$.
The second term penalizes $T_\theta$ from making high-frequency intensity changes to the input image, encapsulating our prior knowledge that skull-stripping and MR bias field correction involve a pixel-wise product with a spatially smooth field.
In this work, we use a total variation penalty~\cite{Osher2005} $L_{TV}$ on the multiplier field $\chi$, which promotes sparsity of spatial gradients in $\chi$. The final objective is:
\begin{equation}
\argmin_\theta 
    L_{rec}(T_\theta(x) \circ \Phi_{aff}, x_{gt}) + \lambda L_{TV}(\chi),
\end{equation}
where $x_{gt}$ is the pre-processed ground truth images, $\lambda \geq 0$ controls the trade-off between the two loss terms, and $\circ$ denotes a spatial transformation.


\subsubsection{Conditioning on $\lambda$.}
Classically, hyperparameters like $\lambda$ are tuned on a held-out validation set - a computationally-intensive task which requires training multiple models corresponding to different values of $\lambda$. 
To avoid this, we condition on $\lambda$ in $f_\theta$ by passing in $\lambda$ as an input to a separate MLP $h_\phi(\lambda)$ (see Fig.~\ref{fig_p}), which generates a $\lambda$-conditional scale and bias for each channel of the decoder layers. 
$h_\phi$ can be interpreted as a hypernetwork~\cite{Ha2016,wang2022computing,Hoopes2021} which generates a conditional scale and bias similar to adaptive instance normalization (AdaIN)~\cite{huang2017arbitrary}.

Specifically, for a given decoder layer with $C$ intermediate feature maps $\{z_1, ..., z_C\}$, $h_\phi(\lambda)$ generates 
the parameters to scale and bias each channel $z_{c}$ such that the the channel values are computed as:
\begin{equation}
    \hat{z}_c = {\alpha_{c}z_{c}}+\beta_{c},
\end{equation}
for $c \in \{1, ..., C\}$.
Here, $\alpha_c$ and $\beta_c$ denote the scale and bias of channel $c$, conditioned on $\lambda$. 
This is repeated for every decoder layer, except the final layer.


\begin{figure}[t!]
\centering
\includegraphics[width=0.8\textwidth,height=4cm]{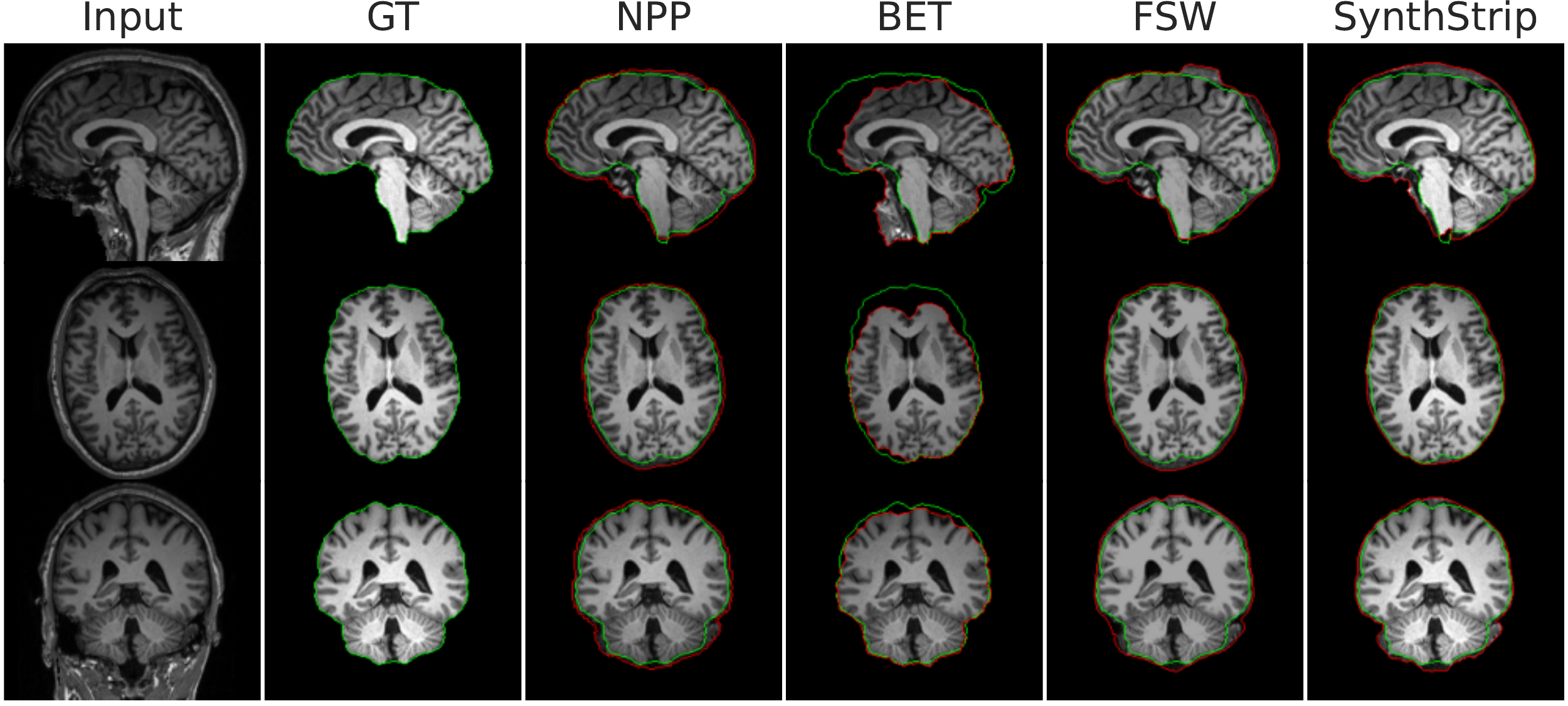}
\caption{Representative slices for skull-stripping. From top to bottom: coronal, axial and sagittal views. Green and red contours depict ground truth and estimated brain masks, respectively.} \label{fig1}
\end{figure}

\section{Experiments}

\subsubsection{Training details.} 
We created a large-scale dataset of 3D T1-weighted (T1w) brain MRI volumes by aggregating 7 datasets: GSP~\cite{Holmes2015}, ADNI~\cite{mueller2005alzheimer}, OASIS~\cite{marcus2010open}, ADHD~\cite{adhd}, ABIDE~\cite{de2017alzheimer}, MCIC~\cite{Gollub2013}, and COBRE~\cite{Aine2017}. 
The whole training set contains 10,083 scans. 
As ground-truth target images, we used FreeSurfer generated skull-stripped, intensity-normalized and affine-registered (to so-called MNI atlas coordinates) images.

We train \modelname~with ADAM~\cite{Kingma2014} and a batch size of 2, for a maximum of 60 epochs. The initial learning rate is 1e-4 and then decreases by half after 30 epochs. 
We use random gamma transformation as a data augmentation technique, with parameter log gamma (-0.3,0.3). We randomly sampled $\lambda$ from a log-uniform distribution on $(-3, 1)$ for each mini-batch.

\subsubsection{Architecture Details.}
$f_\theta$ is a U-Net-style architecture containing an encoder and decoder with skip connections in between. The encoder and decoder have 5 levels and each level consists of 2 consecutive 3D convolutional layers. Specifically, each 3D convolutional layer is followed by an instance normalization layer and LeakyReLU (negative slope of 0.01). In the bottleneck, we use three transformer blocks to enhance the ability of capturing global information~\cite{vaswani2017attention}. Each transformer block contains a self-attention layer and a MLP layer. For the transformer block, we use patch size 1x1x1, 8 attention heads, and an MLP expansion ratio of 1. We perform tokenization by flattening the 3D CNN feature maps into a 1D sequence.

The hypernetwork, $h_\phi$, is a 3-layer MLP with hidden layers 512, 2048 and 496. The STN MLP is composed of a global average pooling layer and a 2-layer MLP with hidden layers of size 256 and 12. The 2-layer MLP contains: linear(256 channels); ReLU; linear(12 channels); Tanh. Note an identity matrix is added to the output affine matrix.

\begin{table}[t]
\parbox[t]{.45\linewidth}{\vspace{0pt}
\centering
\begin{tabular}{|l|*{6}{c|}}
\hline
\textit{Method} &  \textit{SS}  & \textit{IN} & \textit{SN} &\textit{GPU} & \textit{CPU} \\\hline
SynthStrip &  16.5 & - & - & \checkmark&    \\
BET &   9.1& 262.2& -&& \checkmark\\
Freesurfer &  747.2&481.5& 671.6& & \checkmark\\ \hdashline
\modelname~(Ours) &  \multicolumn{3}{|c|}{\textbf{2.94}} & \checkmark&\\
\hline 
  \end{tabular}
  \caption{Supported sub-tasks and average runtime for each method. Skull-stripping (SS), intensity normalization (IN) and spatial normalization (SN). Units are sec, \textbf{bold} is best.}\label{tab_r}
}
\hfill
  \parbox[t]{.45\linewidth}{\vspace{0pt}
\centering
\begin{tabular}{|l|l|l|l|l|}
\hline
 \textit{Method}&\textit{\makecell{Rec \\SSIM}}&\textit{\makecell{Rec \\PSNR}}&\textit{\makecell{Bias \\SSIM}}&\textit{\makecell{Bias \\PSNR}}\\\hline
FSL &  98.5 & 34.2& 92.5 &39.2 \\
FS &  98.9  &35.7 &92.1 &39.3 \\\hdashline
\modelname~&  \textbf{99.1 } & \textbf{36.2 } & \textbf{92.7 } & \textbf{39.4 }\\
\hline
\end{tabular}
  \caption{Performance on intensity normalization. Higher is better, \textbf{bold} is best.}\label{tab_i}
}
\end{table}
\subsubsection{Baselines.}

We chose three popular and widely-used tools, SynthStrip~\cite{hoopes2022synthstrip}, FSL~\cite{Smith2004}, and FreeSurfer~\cite{fischl2012freesurfer}, as baselines. SynthStrip (SS) is a learning-based skull-stripping method, while FSL and  FreeSurfer (FS) is a cross-platform brain processing package containing multiple tools. FSL's Brain Extraction Tool (BET) and FMRIB's Automated Segmentation Tool are for skull stripping and MR bias field correction, respectively. FS uses a watershed method for skull-stripping, a model-based tissue segmentation for intensity normalization and bias field correction.


\subsection{Runtime analyses}

The primary advantage of~\modelname~is runtime. 
As shown in Table \ref{tab_r}, for images with resolution $256 \times 256 \times 256$, \modelname~requires less than 3 sec on a GPU and less than 8 sec on a CPU for all three pre-processing tasks. 
This is in part due to the fact that the multiplier field can be computed at a lower resolution (in our case, on a grid of $128 \times 128 \times 128$). 
The output field is then up-sampled with trilinear interpolation before being multiplied with the input image. 
In contrast, SynthStrip needs 16.5 sec on a GPU for skull stripping. 
FSL's optimized implementation takes about 271.3 sec per image for skull stripping and intensity normalization, whereas FreeSurfer needs more than 10 min.

\begin{figure}[t]
\centering
\includegraphics[width=0.9\textwidth]{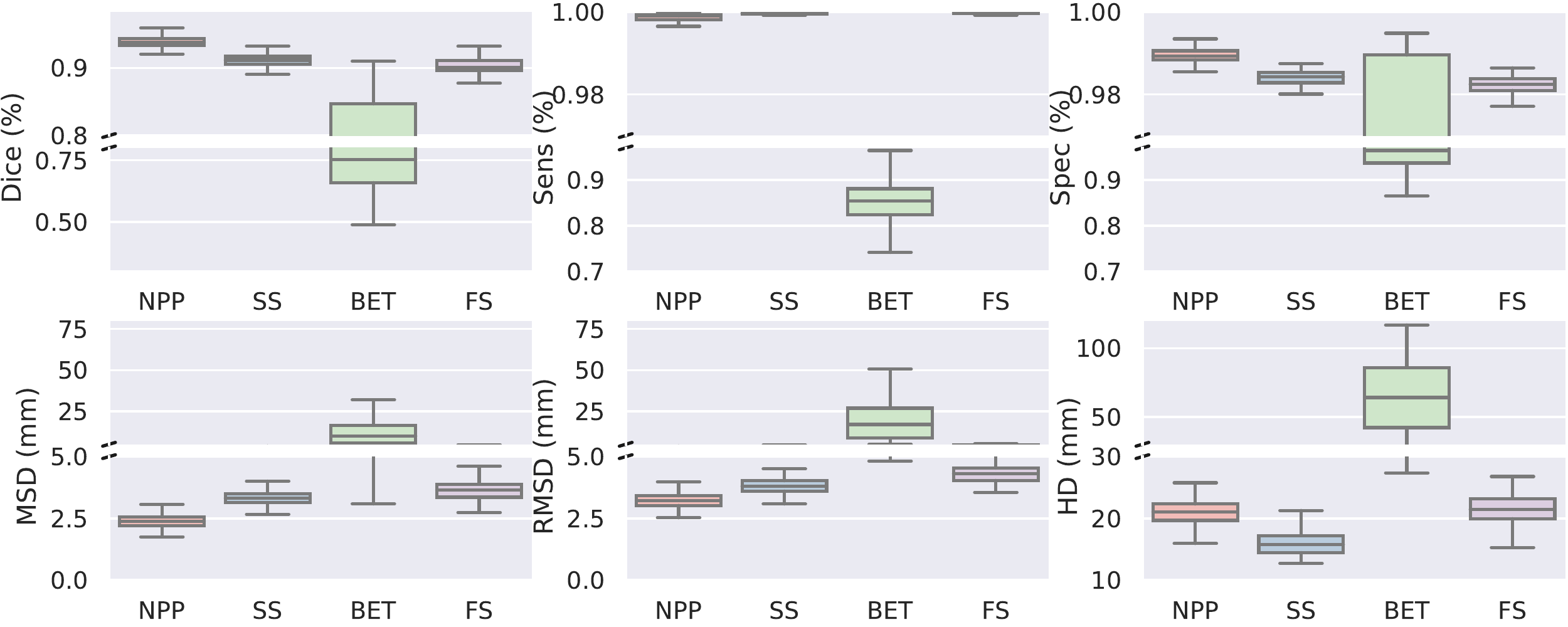}
\caption{Skull-stripping performance on various metrics. (Top) Higher is better. (Bottom) Lower is better.} \label{fig_sr} 
\end{figure}
\subsection{Pre-processing Performance}
We empirically validate the performance of~\modelname~for the three tasks we consider: skull-stripping, intensity normalization, and spatial transformation. 

\subsubsection{Evaluation Datasets.}
For skull-stripping, we evaluate on the Neuralfeedback skull-stripped repository (NFSR)~\cite{Eskildsen2012} dataset. 
NFSR contains 125 manually skull-stripped T1w images from individuals aged 21 to 45, and are diagnosed with a wide range of psychiatric diseases. The definition of the brain mask used in NFSR follows that of FS.  
For intensity normalization, we evaluate on the test set (N=856) from the Human Connectome Project (HCP). 
The HCP dataset includes T1w and T2w brain MRI scans which can be combined to obtain a high quality estimate of the bias field~\cite{Song2022,Glasser2013}.
For spatial normalization, we use T1w MRI scans from the Parkinson's Progression Markers Initiative (PPMI). These images were automatically segmented using Freesurfer into anatomical regions of interest (ROIs)\footnote{
In this work, the following ROIs were used to evaluate performance: brain stem (Bs), thalamus (Th), cerebellum cortex (Cbmlc), cerebellum white matter (Wm), cerebral white matter (Cblw), putamen (Pu), ventral DC (Vt), pallidum (Pa), caudate (Ca), lateral ventricle (LV), and hippocampus (Hi).}~\cite{dalca2018anatomical}.

\subsubsection{Metrics.}
For skull-stripping, we quantify performance using the Dice overlap coefficient (Dice), Sensitivity (Sens), Specificity (Spec), mean surface distance (MSD), residual mean surface distance (RMSD), and Hausdorff distance (HD), as defined elsewhere~\cite{jadon2020survey}.
For intensity normalization, we evaluate the intensity-normalized reconstruction (Rec) and estimated bias image (Bias, which is equal to the multiplier field $\chi$) to the ground truth images, using PSNR and SSIM. 
For spatial normalization, we quantify performance by the Dice score between the spatially transformed segmentations (resampled using the estimated affine transformation) and the probabilistic labels of the target atlas.

\subsubsection{Results.}
Fig~\ref{fig1} and Fig~\ref{fig_sr}~shows skull-stripping performance for all models. 
We observe that the proposed method outperforms all traditional and learning-based baselines on Dice, Spec, and MSD/RMSD. 
Importantly,~\modelname~achieved 93.8\% accuracy and 2.7\% improvement on Dice and 2.39mm MSD.
Especially for MSD,~\modelname~is 28\% better than the second-best method, SynthStrip. 
We further observe that BET commonly fails, which has also been noted in the literature~\cite{ezhilarasan2021automatic}.

Table \ref{tab_i} shows the quantitative results of FSL, FS and~\modelname (see visualization results in Supplementary S1). 
\modelname~outperforms the baselines on all metrics. 
From Table \ref{tab_i}, we see that FreeSurfer's reconstruction is better than BET's, but the bias field estimates are relatively worse. 
We can appreciate this in the figure in Supplementary S1, as we observe that FS's bias field estimate (f) contains too much high-frequency anatomical detail.

\begin{figure}[t]%
    \centering
    \subfloat[\centering ]{{\includegraphics[width=0.5\textwidth,height=4.5cm]{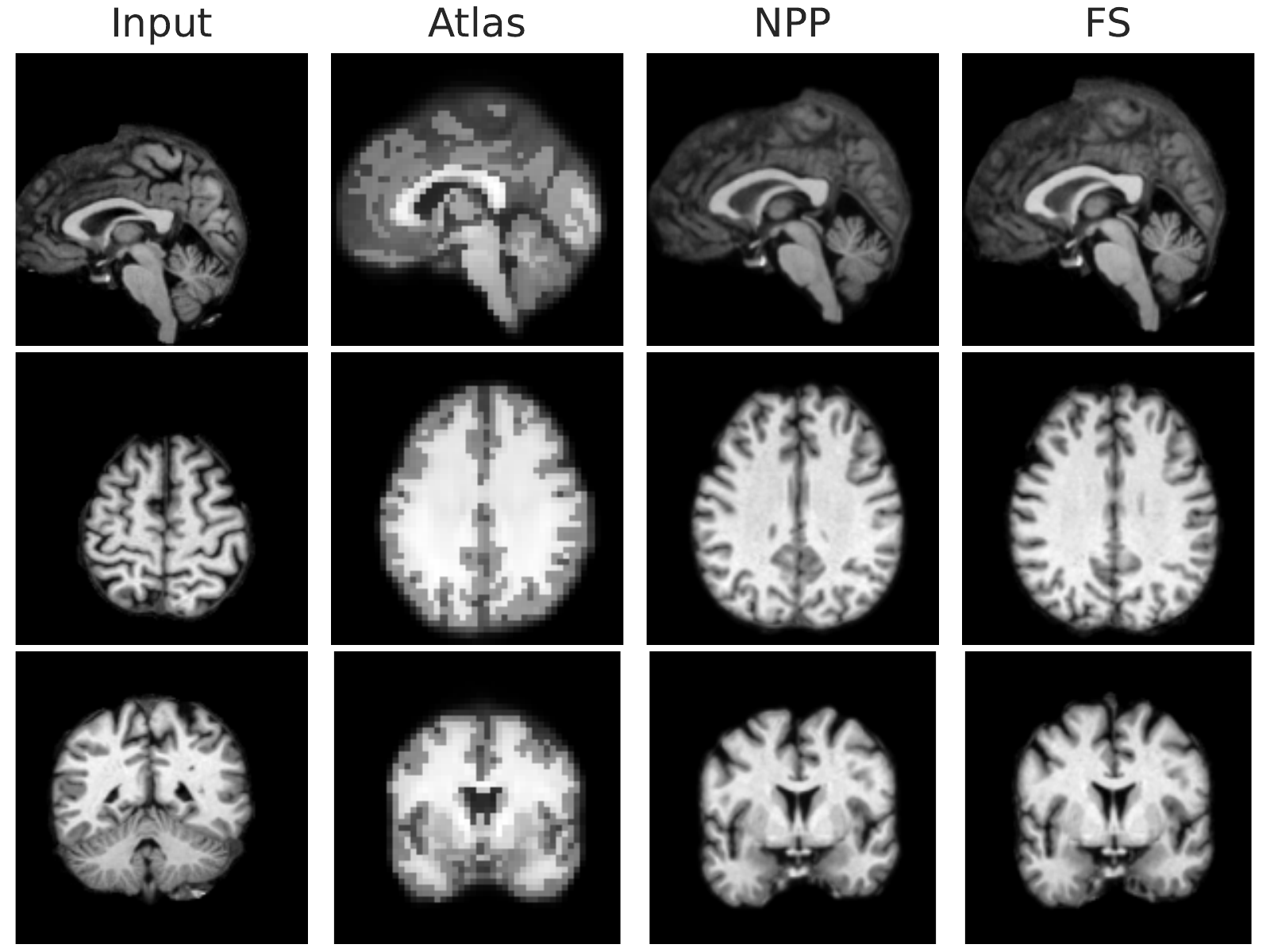}}}%
    \subfloat[\centering ]{{\includegraphics[width=0.47\textwidth,height=5cm]{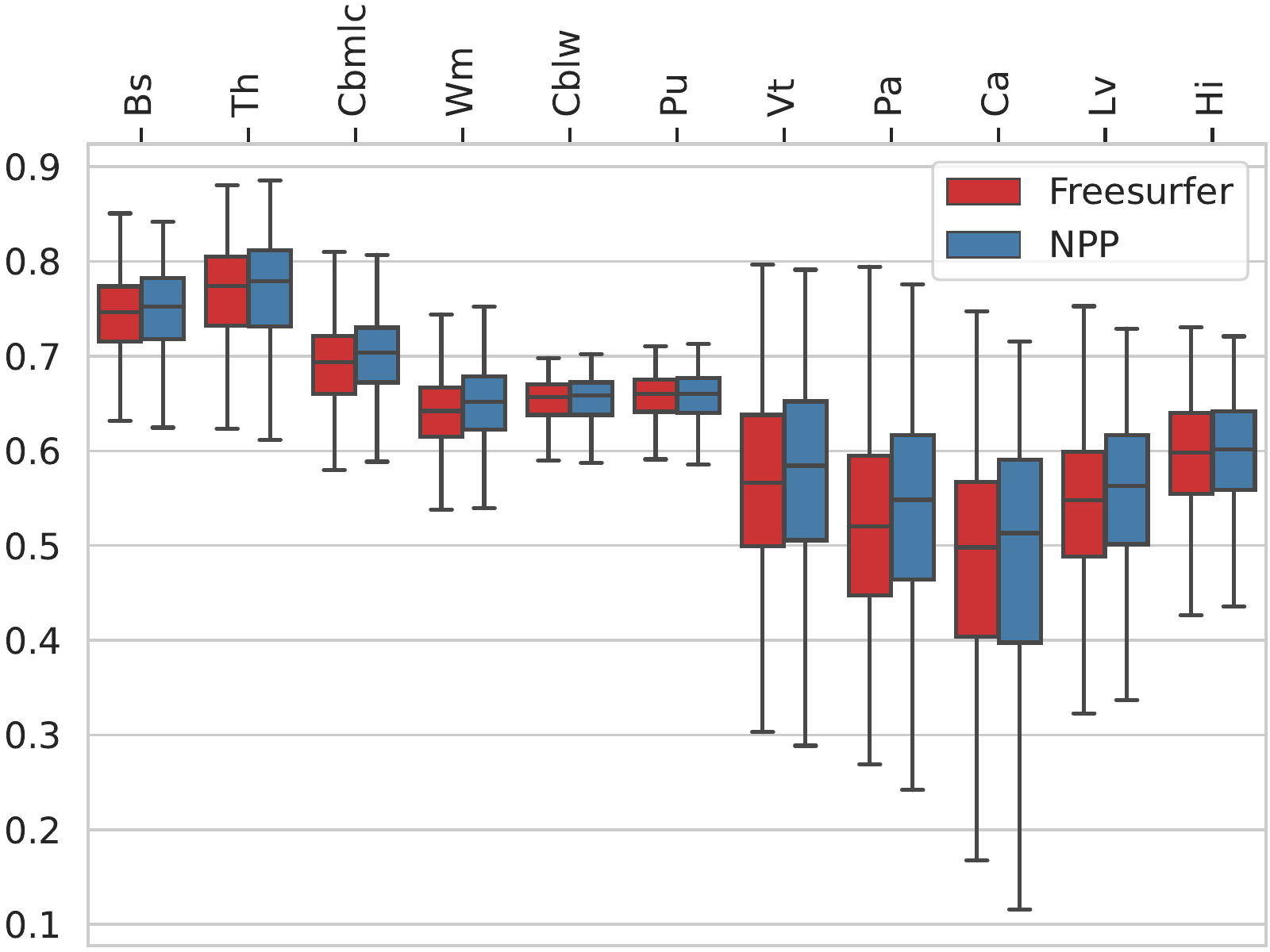}}}%
    \caption{(a) Representative examples for spatial normalization. Rows depict sagittal, axial and coronal view. For each view, from left to right: input image, atlas, \modelname~results, and FreeSurfer results. (b) Boxplots illustrating Dice scores of each anatomical structure for \modelname~and FreeSurfer in the atlas-based registration with the PPMI dataset. We combine left and right hemispheres of the brain into one structure for visualization.}%
    \label{fig_ab}%
\end{figure}

Fig.~\ref{fig_ab} (b) shows boxplots of Dice scores for \modelname~and FreeSurfer, for each ROI. Compared to FS,  
\modelname~achieves consistent improvement on all ROIs measured. Fig.~\ref{fig_ab} (a) shows representative slices for spatial normalization.

\begin{table}[b]
\parbox[t]{.45\linewidth}{\vspace{0pt}
\centering
\caption{Ablation study results of different $\lambda$. \textbf{Bold} is best.}\label{tab_a}

\begin{tabular}{|l|*{6}{c|}}
\hline
\textit{Method} &  \textit{RecSSIM}  & \textit{Bias SSIM}\\\hline
\modelname, $\lambda=10$ & 96.38 $\pm$ 1.29& 96.02 $\pm$ 1.29\\
\modelname, $\lambda=1$ & 99.07 $\pm$ 0.61& 98.09 $\pm$ 0.62\\
\textbf{\modelname, $\lambda=0.1$} & \textbf{99.25 $\pm$ 0.52} & \textbf{98.40 $\pm$ 0.40}\\ 
\modelname, $\lambda=0.01$ & 99.24 $\pm$ 0.51& 98.22 $\pm $0.39\\
\modelname, $\lambda=0.001$ & 99.22 $\pm$ 0.52& 98.13 $\pm$ 0.40\\
\hline
  \end{tabular}
}
\hfill
  \parbox[t]{.45\linewidth}{\vspace{0pt}
\centering
  \caption{Comparison with ablated models. \textbf{Bold} is best.}\label{tab_ad}

\begin{tabular}{|l|l|}
\hline
 Method&\textit{Rec SSIM}\\\hline
Naive U-Net &  84.12 $\pm$ 3.34 \\
U-Net+STN &  84.87$\pm$ 3.04   \\
UMIRGPIT & 84.26  $\pm$ 3.02 
\\\hdashline
\modelname, $\lambda=0.1$ & \textbf{99.25 $\pm$ 0.52}  \\
\hline
\end{tabular}
}
\end{table}

\begin{figure}[t]
\centering
\includegraphics[width=0.9\textwidth]{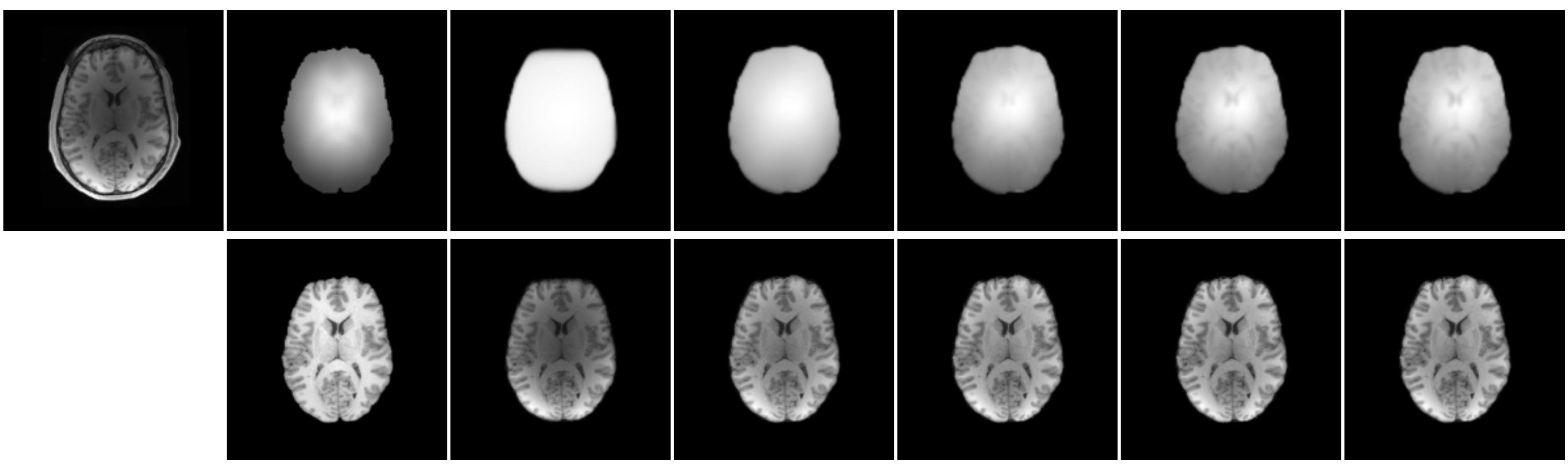}
\caption{(Top) From left to right, raw input image, ground truth bias field, estimated multiplier fields from \modelname~for different values of $\lambda={10,1,0.1,0.01,0.001}$. (Bottom) From left to right, ground truth, outputs from \modelname~for different values of $\lambda$.} \label{fig_a}
\end{figure}

\subsection{Ablation}
As ablations, we compare the specialized architecture of~\modelname~against a naive U-Net trained to solve all three tasks at once. 
Additionally, we implemented a different version of our model where the U-Net directly outputs the skull-stripped and intensity-normalized image, which is in turn re-sampled with the STN. In this version, we did not have the scalar multiplication field and thus our loss function did not include the total variation term. 
We call this version U-Net+STN.
As another alternative, we trained the U-Net+STN architecture via UMIRGPIT \cite{arar2020unsupervised}, which encourages the translation network (U-Net) to be geometry-preserving by alternating the order of the translation and registration. 
We note again that for all these baselines, we used the same architecture as $f_\theta$, but instead of computing the multiplier field $\chi$, $f_\theta$ computes the final intensity-normalized and spatially transformed image directly.
The objective is the reconstruction loss $L_{rec}$. 
All other implementation details were the same as \modelname.
For evaluation, we use the test images from the HCP dataset.
\\

\noindent
\textbf{Results:}
Tables \ref{tab_a} and \ref{tab_ad} lists the SSIM values for the estimated reconstruction and bias fields, for different ablations and \modelname~with a range of $\lambda$ values.
We observe that there is a sweet spot around $\lambda = 0.1$, which underscores the importance of considering different hyperparameter settings and affording the user to optimize this at test time.
All ablation results are poor, supporting the importance of our architectural design.
Fig.~\ref{fig_a} shows some representative results for a range of $\lambda$ values.

\section{Conclusion}

In this paper, we propose a novel neural network approach for brain MRI pre-processing. The proposed model, called \modelname, disentangles geometry-preserving translation mapping (which includes skull stripping and bias field correction) and spatial transformation. 
Our experiments demonstrate that \modelname~can achieve state-of-the-art results for the major tasks of brain MRI pre-processing.

\begin{figure}[t!]
\centering
\includegraphics[width=0.45\textwidth]{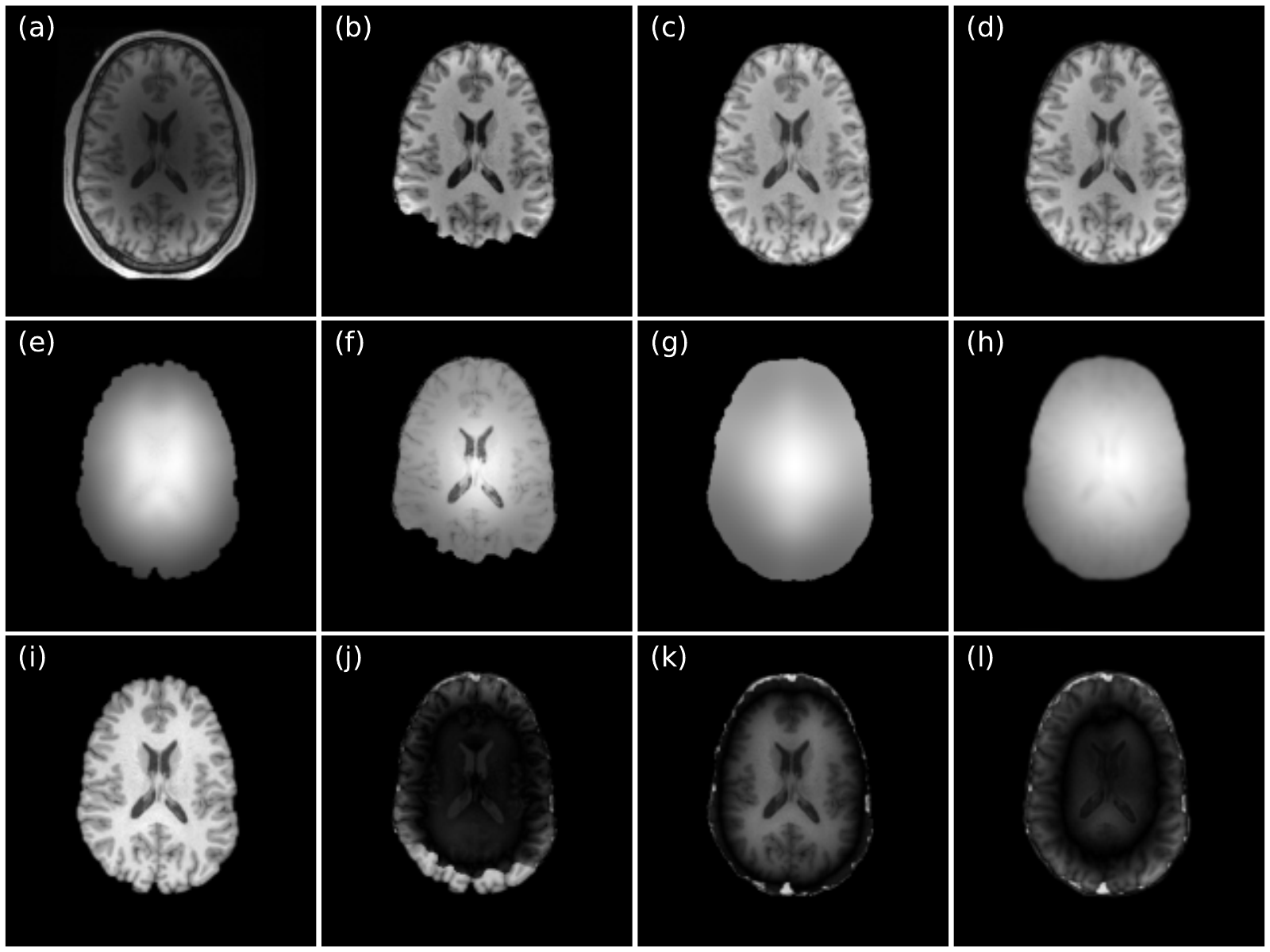}
\caption{Representative slices of intensity normalization performance on HCP T1w brain images. (a) input image; (b-d) intensity normalized outputs from FreeSurfer, FSL, and \modelname~with $\lambda=0.1$; (e) ground truth bias field; (f-h) bias fields estimated by the corresponding methods; (i) ground-truth intensity normalized image; (j-l) prediction error maps corresponding to (b-d).} 
\label{fig_ie}
\end{figure}
\bibliographystyle{splncs04}
\bibliography{miccai}

\end{document}